\documentclass[twocolumn,floatfix,superscriptaddress,aps,pra]{revtex4-2}
\pdfoutput=1
\usepackage[utf8]{inputenc}
\usepackage[english]{babel}
\usepackage[T1]{fontenc}
\usepackage{amsmath}
\usepackage{hyperref}
\usepackage[normalem]{ulem}

\usepackage{graphicx}
\usepackage{amsmath,mathtools}
\usepackage{amssymb}
\usepackage{bm}
\usepackage{color}
\usepackage{bbold}

\usepackage{tikz}
\usepackage{lipsum}

\begin{document}

\title{Ultrafast dynamics of cold Fermi gas after a local quench}

\author{N. V. Gnezdilov}
\email{n.gnezdilov@ufl.edu}
\affiliation{Department of Physics, University of Florida, Gainesville, Florida 32601, USA}
\author{A. I. Pavlov}
\affiliation{The Abdus Salam International Centre for Theoretical Physics (ICTP), Strada Costiera 11, I-34151 Trieste, Italy}
\affiliation{IQMT, Karlsruhe Institute of Technology, 76344 Eggenstein-Leopoldshafen, Germany}
\author{V. Ohanesjan}
\affiliation{Instituut-Lorentz, $\Delta$-ITP, Universiteit Leiden, P.O. Box 9506, 2300 RA Leiden, The Netherlands}
\author{Y. Cheipesh} 
\affiliation{Instituut-Lorentz, Universiteit Leiden, P.O. Box 9506, 2300 RA Leiden, The Netherlands}
\author{K. Schalm}
\affiliation{Instituut-Lorentz, $\Delta$-ITP, Universiteit Leiden, P.O. Box 9506, 2300 RA Leiden, The Netherlands}

\begin{abstract}
We consider non-equilibrium dynamics of two initially independent reservoirs $A$ and $B$ filled with a cold Fermi gas coupled and decoupled by two quantum quenches following one another. We find that the von Neumann entropy production induced by the quench is faster than thermal transport between the reservoirs and defines the short-time dynamics of the system.
We analyze the energy change in the system which adds up the heat transferred between $A$ and $B$ and the work done by the quench to uncouple the reservoirs.
In the case when  $A$ and $B$ interact for a short time, we notice an energy increase in both reservoirs upon decoupling.
This energy gain results from the quench's work and does not depend on the initial temperature imbalance between the reservoirs. We relate the quench's work to the mutual correlations of $A$ and $B$ expressed through their von Neumann entropies. Utilizing this relation, we show that once $A$ and $B$ become coupled, their entropies grow (on a timescale of the Fermi time) faster than the heat flow within the system. This result may provide a track of quantum correlations' generation at finite temperatures which one may probe in ultracold atoms, where we expect the characteristic timescale of correlations' growth to be $\sim 0.1 \,{\rm ms}$. 
\end{abstract}

\maketitle

{\it Introduction.---}  Experimental techniques in ultracold quantum gases have greatly advanced in recent years, providing a vigorous control over transport phenomena \cite{Brantut2012Conduction, Stadler2012Observing, Brantut2013Thermoelectric, Krinner2013Superfluidity, Krinner2014Observation, Valtolina2015Josephson, Husmann2015Connecting, Chien2015Quantum, Grenier2016Thermoelectric, Krinner2017Two, Burchianti2018Connecting, Husmann2018Breakdown, Kwon2020Strongly, Luick2020Ideal, DelPace2021Tunneling, Zhou2022Observation}. 
In contrast to their electronic counterparts, the reservoirs formed out of trapped cold Fermi gas are well-isolated from the outer environment and allow for highly tunable interaction strength and disorder. This level of adjustment makes the ultracold atomic systems particularly attractive to probe non-equilibrium dynamics of quantum many-body systems in transport observables. 

Due to the atomic nature of carriers, the characteristic timescales of the tunneling phenomena differ by many orders from the electron transport. The shortest timescale relevant for transport in a Fermi system is the Fermi time $\tau_{\rm F}\!\sim\!1/\varepsilon_{\rm F}$ -- the time a particle travels a distance comparable to the Fermi wavelength, where $\varepsilon_{\rm F}$ is the Fermi energy. In turn, the transport measurements are performed on a timescale much longer than the Fermi time \cite{Krinner2014Observation}. Indeed, for a quantum point contact, it takes $\sim  10 \, \tau_{\rm F}$ to form a steady flow pattern after inducing the potential difference within a system \cite{Beria2013Quantum, Krinner2014Observation}. Whereas in an electronic setup, this falls into the category of ultrafast processes, being in the femtosecond range, in ultracold atoms, the Fermi time is on the order of $0.1 \,{\rm ms}$. The magnitude of the timescale difference allows one to study the processes specific to ultrafast physics in a moderate millisecond time frame \cite{Cetina2016Ultrafast}. 

A natural way to study the early-time evolution of a many-body system is via quantum quench. A quantum quench drives the system out of equilibrium by an explicit change of a system's Hamiltonian parameters \cite{Polkovnikov2011}, e.g., turning on or off the interaction between the subparts of a composite system. Achievable in highly controlled cold atomic platforms, the post-quench dynamics provide significant insights into the keystone concepts of many-body physics such as entanglement, ergodicity, and thermalization \cite{Klich2009Quantum, Fradkin2009PRB, Cardy2011Measuring, Polkovnikov2011, Abanin2012Measuring, Islam2015Measuring, Alba2017Entanglement, Serbyn2021Quantum}.

Conventionally, one considers the early-time dynamics of a many-body system after the quantum quench in a nearly adiabatic regime. In this case, the quench turns on slowly compared to the characteristic timescale of the problem \cite{Polkovnikov2011}. Instead, interested in a system's dynamics on a timescale comparable to the Fermi time, we focus on a local quench that instantly changes the Hamiltonian of the system. 

The evolution of a quantum system after the local quench is known to pave the way towards measuring entanglement entropy \cite{Klich2009Quantum, Fradkin2009PRB, Cardy2011Measuring, Abanin2012Measuring, Islam2015Measuring}. The entanglement entropy is a measure of non-classical correlations in composite quantum systems commonly defined as the von Neumann entropy of a subpart of a total system which, in turn, is described by a pure state \cite{Nielsen2009Quantum}.
At zero temperature, the local quench connecting the two subspaces of a bipartite system generates entanglement entropy measurable in the particle density fluctuations in free fermion and fractional quantum Hall systems \cite{Klich2009Quantum,Fradkin2009PRB}. 
The generalization of Refs. \cite{Klich2009Quantum, Fradkin2009PRB} to finite temperature is not straightforward since entanglement and thermal contributions to the von Neumann entropy are nonadditive. However, if the temperature is low enough, one expects an instant coupling of two tanks of cold Fermi gas prepared at different temperatures to generate quantum correlations between them.
Then, a question arises:  What is the characteristic timescale of correlation generation initiated by the quench coupling, and how does this timescale compare to the one of the heat flow due to the initial thermal imbalance? 

In this paper, we show that correlation generation induced by a local quench is an ultrafast process. 
To track the correlation generation, 
we consider two reservoirs $A$ and $B$ filled with cold non-interacting Fermi gas sequentially coupled and decoupled by the two local quenches. We evaluate the von Neumann entropy production in the reservoirs by relating it to the energy change of $A$ and $B$ after decoupling. Then we compare the resulting entropy dynamics to the heat current within the system. We find that the entropy production occurs on a timescale of $\tau_{\rm F}$, which is considerably faster than the thermal transport.

{\it Thermal state driven out of equilibrium.---} 
We begin with two systems $A$ and $B$, each initially prepared in a thermal state, that are instantaneously coupled by an interaction term $V_{AB}$. The generic Hamiltonian is
\begin{align}
    H(t) = H_A + H_B + g(t) V_{AB}, \label{H}
\end{align}
where the function $g(t) = \theta(t) - \theta(t-\tau)$ defines a quench protocol that couples $A$ to $B$ at time $t=0$ and disconnects them at $t = \tau$.

The initial state of the full system is given by the product of two thermal density matrices 
\begin{align}
   \rho_0 =& \rho_A \otimes \rho_B, \label{rho_0}\\ 
   \rho_\alpha =& Z_\alpha^{-1}\sum_{n_\alpha} e^{-\frac{E_{n_\alpha}}{T_\alpha}} |n_\alpha\rangle \langle n_\alpha| = e^{\frac{\mathcal{F}_\alpha-H_\alpha}{T_\alpha}}, 
   \label{rho_thermal}
\end{align}
Here 
$|n_\alpha\rangle$ is an eigenstate of the Hamiltonian $H_\alpha$ with energy $E_{n_\alpha}$, $T_\alpha$ is the initial temperature, $\mathcal{F}_\alpha = - T_\alpha \ln Z_\alpha$ is the thermal free energy, and $Z_\alpha = \mathrm{Tr}_\alpha e^{-H_\alpha/T_\alpha}$ is the partition function for $\alpha = A, B$ \footnote{We use the units $\hbar = k_{\rm B} = 1$}. 

Once the two systems are coupled they become correlated. A natural measure to study the correlations between $A$ and $B$ is the von Neumann entropy.
The von Neumann entropy for system $A$ is 
\begin{align}
    S_{\rm vN}(t) = - \mathrm{Tr}_A \, \rho_A(t) \ln \rho_A(t), \label{S_VN}
\end{align}
where $\rho_A(t) = \mathrm{Tr}_B \, U(t) \rho_0 U^\dag(t)$ is reduced density matrix and $U(t) = {\rm \hat{T}} \exp(-i \int_0^t dt' H(t'))$ is the time-ordered evolution operator.

Let us introduce the relative entropy, which is often used in both quantum information processing \cite{Nielsen2009Quantum} and quantum thermodynamics \cite{Deffner2019Quantum} to distinguish between two quantum states and as a measure of irreversibility of a thermodynamic process \cite{Funo2013Thermodynamic}. 
For our purpose, we define the relative entropy between the evolved state $\rho_A(t)$ of the system $A$ from its initial thermal state $\rho_A$: 
\begin{align} 
    S\left(\rho_A(t) || \rho_A \right) = \mathrm{Tr}_A \, \rho_A(t) \left( \ln \rho_A(t) -  \ln \rho_A \right) \geq 0. \label{S_relative}
\end{align}
 
Using that the initial state of $A$ is a thermal state at temperature $T_A$, we relate the expectation value of the Hamiltonian $H_A$ to the combination of the von Neumann entropy (\ref{S_VN}) and the relative entropy (\ref{S_relative}) \footnote{We use that $H_A$ can be expressed as $H_A = \mathcal{F}_A - T_A \ln \rho_A$.}: $\mathrm{Tr}_A \, \rho_A(t) H_A = \mathcal{F}_A - T_A \mathrm{Tr}_A \, \rho_A(t) \ln \rho_A = \mathcal{F}_A + T_A\left(S_{\rm vN}(t) + S(\rho_A(t)||\rho_A)\right)$. Subtracting the initial energy value $\mathrm{Tr}_A \, \rho_A H_A = \mathcal{F}_A + T_A S_{\rm vN}(0)$ from $\mathrm{Tr}_A \, \rho_A(t) H_A$, we get
\begin{align}
    \Delta E_A(t) = T_A\left(\Delta S_{\rm vN}(t) + S(\rho_A(t)||\rho_A)\right), \label{I_law}
\end{align}
where $\Delta E_A(t) = \mathrm{Tr}_A \, \rho_A(t) H_A - \mathrm{Tr}_A \, \rho_A H_A$ and $\Delta S_{\rm vN}(t) = S_{\rm vN}(t) - S_{\rm vN}(0)$.
The relation (\ref{I_law}) is the first law of thermodynamics for a subpart of a composite quantum system driven from its initial thermal state \footnote{In quantum thermodynamics \cite{Deffner2019Quantum}, the equation (\ref{I_law}) is often written as $\Delta E_A = T_A \Delta S_{\rm vN} + \Delta\tilde{\mathcal{F}}_A$, where $\tilde{\mathcal{F}}_A(t) = \mathcal{F}_A + T_A S(\rho_A(t)||\rho_A)$ is the information free energy -- a generalization of thermal free energy for out-of-equilibrium processes.}.
A thermodynamic standpoint on the evolution of a quantum system enables one to characterize the irreversibility of a dynamical process \cite{Dorner2012Emergent} and the emergence of decoherence \cite{Popovic2021Thermodynamics} and to relate multipartite quantum correlations to extractable work \cite{Alicki2013Batteries, Hovhannisyan2013Batteries, Rossini2020Quantum}.

Equation (\ref{I_law}) is most appropriately seen as a thermodynamic statement. It establishes the energy-to-entropy balance after the process is over, i.e. the two systems are decoupled. Indeed, one can not completely isolate the two systems from each other when they are coupled and determine the actual energy shift in $A$ or $B$.
As such, we shall understand $\Delta E_{A/B}$ as the energy change in the system after decoupling at $t=\tau$. 

At zero temperature, turning on the interaction between the subparts of a composite quantum system may induce quantum correlations that increase the von Neumann entropy of each subpart \cite{Klich2009Quantum, Fradkin2009PRB}. At finite temperature, when $A$ and $B$ are decoupled the energies of both systems will change by $\Delta E_{A/B} \geq T_{A/B} \Delta S_{\rm vN}$, where we used that the relative entropy is non-negative \cite{Nielsen2009Quantum}.  So if the change of the von Neumann entropy is positive, the energy change in the system is also inevitably positive. 
In particular, such an energy increase may be relevant for quantum technology applications, e.g., quantum digital cooling, where a quantum system is brought to the low-energy state by a coupling/decoupling protocol with a cool-bath \cite{Polla2021Quantum}.
Utilizing the free fermions example, we demonstrate that at low temperatures the von Neumann entropy increases under the fast decoupling condition $\tau \lesssim \tau_{\rm F}$. 

{\it The case study: free fermions.---} In free fermion systems, quantum correlations in the Fermi sea are well studied \cite{Beenakker2003Proposal,Klich2006Lower,Beenakker2006ElectronHole}, including the generation of quantum correlations after a local quench where the entanglement entropy is related to the particle number fluctuations \cite{Klich2009Quantum}. At the same time, free-particle motion defines transport properties in ultracold Fermi gas \cite{Krinner2014Observation}. Hence, we proceed with a free fermions model to compare the characteristic timescales for thermal transport and the entropy production induced by a local quench.

Consider for systems $A$ and $B$ two two-dimensional reservoirs with spinless free fermions. 
The Hamiltonian (\ref{H}) reads
\begin{align}
    H_A =& \sum_{\bf p} \xi_{\bf p} a^\dag_{\bf p} a_{\bf p}, \label{HA} \quad
    H_B = \sum_{\bf p} \xi_{\bf p} b^\dag_{\bf p} b_{\bf p},\\
    V_{AB} =& \lambda  a^\dag({\bf r}=0) b({\bf r}=0) + h.c. ,
\end{align}
where $A$ and $B$ are coupled locally in space at ${\bf r}=0$. Here $a, a^\dag$ and $b, b^\dag$ are the fermionic operators in the reservoirs $A$ and $B$, ${\bf p}$ is the momentum, $\xi_{\bf p}$ is the corresponding dispersion, and $\lambda$ is the coupling constant. The size of each reservoir is $V$.  Both reservoirs are at equal chemical potential $\mu \simeq \varepsilon_{\rm F}$.  Note that $[H_\alpha, V_{AB}]\neq 0$.

\begin{figure}[t!!]
 \center \vspace{-0.92cm}
\includegraphics[width=0.928\linewidth]{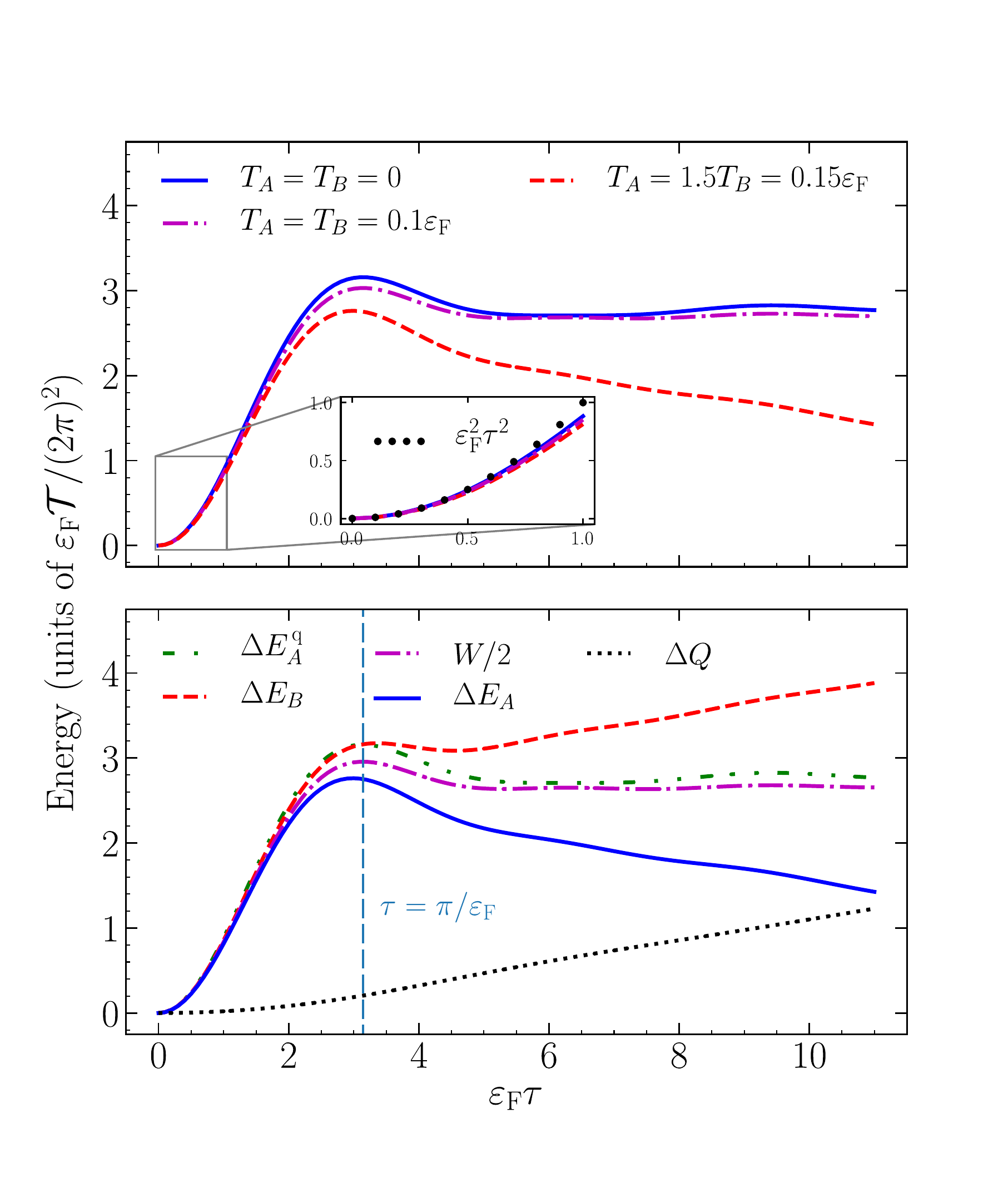} \vspace{-0.6cm}
\caption{\small \label{fig:E_2} (Top) Energy increment in the reservoir $A$, $\Delta E_A$, due to quench-coupling with $B$ as a function of time calculated from Eq.~(\ref{dHdt}). 
The inset demonstrates that the approximation of $\tau \ll 1/\varepsilon_{\rm F}$ in Eq.~(\ref{dE_T0}) accurately describes the energy increment up to $\tau\!\sim\!\ 1/\varepsilon_{\rm F}$ for a given initial temperature imbalance. 
(Bottom) Energy balance in the system upon decoupling at $t=\tau$ for $T_A = 0.15 \varepsilon_{\rm F}$ and $T_B = 0.1 \varepsilon_{\rm F}$. $\Delta E_A^{\, \rm q}$ is the energy change at zero temperature (\ref{dE_T0}), $\Delta E_B$ is the energy change in the initially colder system, $\Delta E_A$ is the energy change in the initially hotter system, $W$ is the work done by the quench upon decoupling, and $\Delta Q$ is the heat transferred from $A$ to $B$. The vertical line marks the maximum of $\Delta E_A^{\, \rm q}$. The figures for a different set of initial temperatures are presented in the Appendix \ref{app:dQ}.
}
\end{figure}


We begin our analysis with the energy transfer in the system.
To determine the overall energy shift in the reservoir $A$, we compute the energy flux $ \frac{d\langle H_A \rangle}{dt} = i g(t) \langle [V_{AB}, H_A] \rangle  = - i g(t) V^{-1} \sum_{{\bf p}{\bf p}'}\xi_{\bf p} ( \lambda \langle a^\dag_{\bf p} b_{{\bf p}'}\rangle -h.c. )$
within time-dependent perturbation theory in $\lambda$ \footnote{See Appendix \ref{app:FGR}}. 
Consequently, in the lowest order, we obtain the Fermi golden rule formula for the energy shift
\begin{align} \nonumber
    \Delta E_A =& -\frac{\cal T}{(2\pi)^2} \!\! 
    \int\limits_{-\varepsilon_{\rm F}}^{\varepsilon_{\rm F}} \!\! d\omega  d\omega' \, \omega \, \frac{\sin^2(\delta\omega\tau/2)}{ (\delta\omega/2)^2}  \\ & \times (n_A^{(0)}(\omega) - n^{(0)}_B(\omega')), \label{dHdt}
\end{align}
where ${\cal T} = (2\pi)^2 \nu_A \nu_B |\lambda|^2$ is the transmission coefficient \cite{Datta1995Electronic}, $\delta\omega = \omega -\omega'$, 
and $n_\alpha^{(0)}(\omega) = (e^{\omega/T_\alpha}+1)^{-1}$ are the initial occupation numbers. 
In the above, we introduced the density of states $\nu_\alpha = V^{-1}\sum_{\bf p}\delta(\omega-\xi_{\bf p}) = p_{\rm F}^2/(4\pi \varepsilon_{\rm F})$ and
replaced sums over momenta with integrals over energy \footnote{The density of states (DoS) in ultracold atomic gases is usually not constant due to the inhomogeneity of the trap potential. In the Appendix \ref{app:trap}, we demonstrate that consideration of a harmonically trapped non-interacting Fermi gas with DoS $\propto \omega^2$ \cite{Brantut2013Thermoelectric, Grenier2016Thermoelectric} does not qualitatively change the energy curves}. 
Here $\varepsilon_{\rm F}$ is the Fermi energy in the reservoir that we use as the UV cutoff for the energy integrals and $p_{\rm F}$ is the Fermi momentum.

Let us consider $A$ and $B$ at zero temperature. Turning on the coupling entangles the states in the reservoirs and, thus, generates entanglement entropy between previously disconnected systems \cite{Klich2009Quantum}. Alongside, energy measurements are known to exhibit entanglement properties in a quantum system \cite{Jordan2004Entanglement}.
Hence, we 
investigate whether the energy of the reservoirs remains unchanged after decoupling. 


At zero temperature the distribution function is $n_\alpha^{(0)}(\omega)=\theta(-\omega)$ for both reservoirs. Substituting the unit-step distribution functions into Eq.~(\ref{dHdt}) and evaluating the energy integrals, we derive the energy shift in the reservoir $A$: 
\begin{align}
    \Delta E_A^{\,\rm q} = \frac{\cal T}{2\pi} \, \varepsilon_{\rm F} \! \int\limits_0^{\varepsilon_{\rm F} \tau} \! \frac{d\zeta}{\pi} \, \sin\zeta \,  \frac{\sin^2\left(\zeta/2\right)}{\left(\zeta/2\right)^2}. \label{dE_T0}
\end{align}
The second reservoir acquires equal energy increment $\Delta E_B^{\,\rm q}=\Delta E_A^{\,\rm q}$. 
As shown in Fig.~\ref{fig:E_2} (Top) (solid blue curve), the energy of the reservoir increases in the absence of temperature or particle imbalances. 
For times $\tau \ll 1/\varepsilon_{\rm F}$, the energy grows quadratic in time: $\Delta E_A^{\,\rm q} \simeq {\cal T}/(2\pi)^2  \varepsilon_{\rm F}^3  \tau^2$.
The superscript ``${\rm q}$'' punctuates a quantum origin of the effect obtained within zero-temperature quantum-mechanical perturbation theory.

Now suppose that the reservoirs $A$ and $B$ are prepared at low temperatures ($T_A, T_B \ll \varepsilon_{\rm F}$) and consider a cooling protocol for the reservoir $A$: $T_A > T_B$. The temperature imbalance between the reservoirs inevitably leads to heat transport. The heat current across the tunneling contact is ${\cal I}_{T} = - \frac{d}{dt} \frac{1}{2} \sum_{\bf p} \xi_{\bf p}(\langle  a^\dag_{\bf p} a_{\bf p} \rangle - \langle  b^\dag_{\bf p} b_{\bf p}\rangle)$. 
We evaluate the overall heat transmitted from $A$ to $B$ by the moment of decoupling as $\Delta Q = \int dt \, {\cal I}_T$, leading to
\begin{align} 
    \Delta Q =& \frac{1}{2}(\Delta E_B - \Delta E_A),
    \label{dQ}
\end{align}
and  focus on a short-time limit $\tau\!\sim\!1/\varepsilon_{\rm F}\!\ll\!1/{\rm max}(T_A,T_B)$. The heat transfer in Eq.~(\ref{dQ}) accounts for the relative energy flux between the reservoirs to exclude the external contribution due to the explicit time dependence of the Hamiltonian (\ref{H}). 


We plot 
$\Delta Q$ computed from Eqs.~(\ref{dQ},\ref{dHdt}) in Fig.~\ref{fig:E_2} (Bottom) (dotted black curve). Comparing the heat to the energy curves in Fig.~\ref{fig:E_2} (Bottom), one notices that for short $\tau$, the heat transfer is considerably slower than the energy increment. Pushing the short-time limit to the extreme, $\varepsilon_{\rm F} \tau \ll 1$, we find that the heat transfer is suppressed by the temperature-dependent coefficient if compared to the energy change in the same regime. Estimating the finite temperature corrections to Eq.~(\ref{dE_T0}) using the Sommerfeld expansion \footnote{See Appendix \ref{app:E_T}}, we find  
$\Delta Q  \propto \pi^2 ( T_A^2 - T_B^2 ) \varepsilon_{\rm F} \tau^2/6$, so that the ratio $\Delta Q/\Delta E_A^{\,\rm q} = \pi^2 (T_A^2-T_B^2)/(6\varepsilon_{\rm F}^2)$ vanishes in the low-temperature limit. 

As shown in Fig.~\ref{fig:E_2}, the energy increment in both reservoirs does not depend on temperature up to $\tau\!\sim\!1/\varepsilon_{\rm F} $ and is well described by the quantum contribution (\ref{dE_T0}) ($T_A=T_B=0$). 
The refrigerated system starts showing energy decrease (cooling down) near $\tau\!\sim\!\pi/\varepsilon_{\rm F} $, which corresponds to the maximum of the zero-temperature energy curve (\ref{dE_T0}). Accordingly, the heat contribution to energy increases as $\tau$ approaches the inverse temperature \footnote{Being interested in the short-time dynamics after the local quench, we disregard the late-time equilibration of the reservoirs. Refs. \cite{Almheiri2019Universal,Zhang2019Evaporation,Ohanesjan2022Energy} access the equilibration regime for the interacting systems, using exact diagonalization \cite{Almheiri2019Universal,Ohanesjan2022Energy} and a numerical solution of the Kadanoff-Baym equations \cite{Zhang2019Evaporation}, indicating the energy curves decay exponentially to an equilibrium value.}.

The heat transfer lowers the energy in the reservoir $A$. Thus, the energy gain originates from external work. Indeed, the generic form of the Hamiltonian (\ref{H}) implies that energy can be added to or subtracted from the total system when turning on or off the interaction between $A$ and $B$: $d \langle H\rangle/d t = \delta(t) \langle V_{AB}(0) \rangle - \delta(t-\tau) \langle V_{AB}(\tau) \rangle$. Evaluating the expectation value of the coupling term as we did for the energy (\ref{dHdt}), we find $\langle V_{AB}(\tau) \rangle = - \Delta E_A - \Delta E_B$ and consequently $\langle V_{AB}(0) \rangle = 0$, meaning that the first quench does not transfer energy in or out of the system. Hence, the total energy of $\Delta E_A +\Delta E_B$ is added to the system at $t = \tau$,
\begin{align}
    \frac{d \langle H \rangle}{d t} =  \delta(t-\tau) \, W, \label{energy_flux_2}
\end{align}
as a work $W = \Delta E_A +\Delta E_B$ done by the quench to decouple the two reservoirs. Alternatively, one can think of the energy $\Delta E_A +\Delta E_B$ as the binding energy of $A$ and $B$ -- the energy required to decouple the reservoirs. 

The work $W$ does not depend on the thermal gradient in the system, as shown in Fig.~\ref{fig:E_2} (Bottom) (dash-dotted magenta curve compared to the solid blue and dashed red curves). Furthermore, on a timescale of $\tau \lesssim 1/\varepsilon_{\rm F}$, $W$ defines the energy increment in both reservoirs, whereas the latter is given by the zero temperature result (\ref{dE_T0}).
Combining Eqs.~(\ref{dQ},\ref{energy_flux_2}), we find for $\tau \lesssim 1/\varepsilon_{\rm F}$ and $T_A, T_B \ll \varepsilon_{\rm F}$ that
\begin{align}
    \Delta E_{A/B} = \mp \Delta Q  + \frac{W}{2} \simeq \frac{W}{2}, \label{E_to_W}
\end{align}
where we neglect the heat transfer compared to the work contribution to the reservoir's energy. In the above, the ``$-$'' sign is taken for the reservoir $A$ and the ``$+$'' sign for the reservoir $B$.

Having the energy dynamics set, we proceed to the entropy analysis to account for the correlations' generation. 
From here on, we consider the reservoirs at equal temperatures $T_A = T_B = T \ll \varepsilon_{\rm F}$ since the short-time dynamics ($\tau \lesssim 1/\varepsilon_{\rm F}$) is not affected by temperature imbalance.

Once we decouple the system at $t = \tau$, while it remains well isolated from the outer environment, each reservoir pursues unitary evolution. We assume that the reservoirs are independent if observed much later after decoupling \footnote{In this case, the reservoirs are described by the Hamiltonians $H_A$ and $H_B$ correspondingly.}. In this case, their von Neumann entropies equal the entropy of a Fermi gas expressed in their occupation numbers \cite{Landau1980Statistical} and coincides with the diagonal entropy known for contributing to the energy change in an out-of-equilibrium process \cite{Polkovnikov2011Microscopic}.  For the reservoir $A$, the entropy is $S_{\rm vN}(t) = -\sum_{\bf p} ( \langle a_{\bf p}^\dag a_{\bf p}\rangle \ln  \langle a_{\bf p}^\dag a_{\bf p}\rangle + (1 - \langle a_{\bf p}^\dag a_{\bf p}\rangle ) \ln (1 -  \langle a_{\bf p}^\dag a_{\bf p}\rangle ) )$. Since the post-decoupling unitary evolution of each reservoir implies the von Neumann entropy conservation, we have $S_{\rm vN}(t=\tau) = S_{\rm vN}(t\gg \tau)$.

To evaluate the entropy we begin with formally expanding it in occupation numbers: 
\begin{align}  \nonumber
    S_{\rm vN}(\tau) =& S_{\rm vN}^{(0)} - \sum_{\bf p} ( n^{(1)}_{\bf p}(\tau) + n^{(2)}_{\bf p}(\tau) + \ldots )  \ln  \frac{n^{(0)}_{\bf p}}{1 -  n^{(0)}_{\bf p}} \\ &- \sum_{\bf p} \frac{n^{(1)}_{\bf p}(\tau)^2}{2n^{(0)}_{\bf p} (1 -  n^{(0)}_{\bf p} )} + \ldots ,\label{S_A_pert}
\end{align}
where $n^{(m)}_{\bf p}\!(\tau)$ are perturbative corrections to equilibrium occupation numbers $n_{\bf p}^{(0)}=(e^{\xi_{\bf p}/T}\!\!+\!1)^{-1}$ with the superscript $m$ marking
the order in ${\cal T}$ \footnote{We omit the reservoir index since  $n_A = n_B$ in absence of the initial particle and temperature imbalances.}. 
The first term in Eq.~(\ref{S_A_pert}) is the initial entropy of the reservoir $S^{(0)}_{\rm vN} = N (\pi^2 /3) (T/\varepsilon_{\rm F})$, where $N = V p_{\rm F}^2/(4\pi)$ is the number of particles
\footnote{We evaluate the initial entropy of the reservoir using the density of the fermion occupation number, see e.g. Ref. \cite{Klich2009Quantum}, $\mu(z) = \sum_{\bf p}\delta(z - n^{(0)}_{\bf p} ) =  N T/\varepsilon_{\rm F}/(z(1-z))$. 
Then the entropy is $S^{(0)}_{\rm vN} =  -\int_0^1 dz \, \mu(z) ( z \ln z + (1-z)\ln (1-z) ) = N (\pi^2 /3) (T/\varepsilon_{\rm F})$.}. The second term in Eq.~(\ref{S_A_pert}) is the overall energy increment $\Delta E_A$ divided by the initial temperature $T$, where we used $n^{(0)}_{\bf p}/(1-n^{(0)}_{\bf p}) = e^{-\xi_{\bf p}/T}$. 
We compare Eq.~(\ref{S_A_pert}) to Eq.~(\ref{I_law}) and combine the remaining terms in Eq.~(\ref{S_A_pert}) into the relative entropy taken with a minus sign:
\begin{align}
    S(\rho_A(\tau)||\rho_A) = \sum_{\bf p} \frac{n^{(1)}_{\bf p}(\tau)^2}{2n^{(0)}_{\bf p} (1 -  n^{(0)}_{\bf p} )} - \ldots. \label{S_rel_pert}
\end{align}

In contrast to the energy increment obtained within quantum-mechanical perturbation theory and most prominent at zero temperature, the entropy computation explicitly requires $T \neq 0$. Indeed, taking the lower energy bound in Eq.~(\ref{I_law}) leads to the entropy divergence $\Delta S_{\rm vN} = \Delta E_A/T$ at $T \to 0$ since the primary contribution to energy (\ref{dE_T0}) does not depend on temperature. Furthermore, the relative entropy may also diverge at low temperatures due to the Fermi functions in the denominator in Eq.~(\ref{S_rel_pert}). Thus, combining quantum mechanical perturbation theory with non-equilibrium thermodynamics requires a lower bound on temperature $T^*$.

\begin{figure}[t!!]
\center \vspace{-0.08cm}
\includegraphics[width=0.928\linewidth]{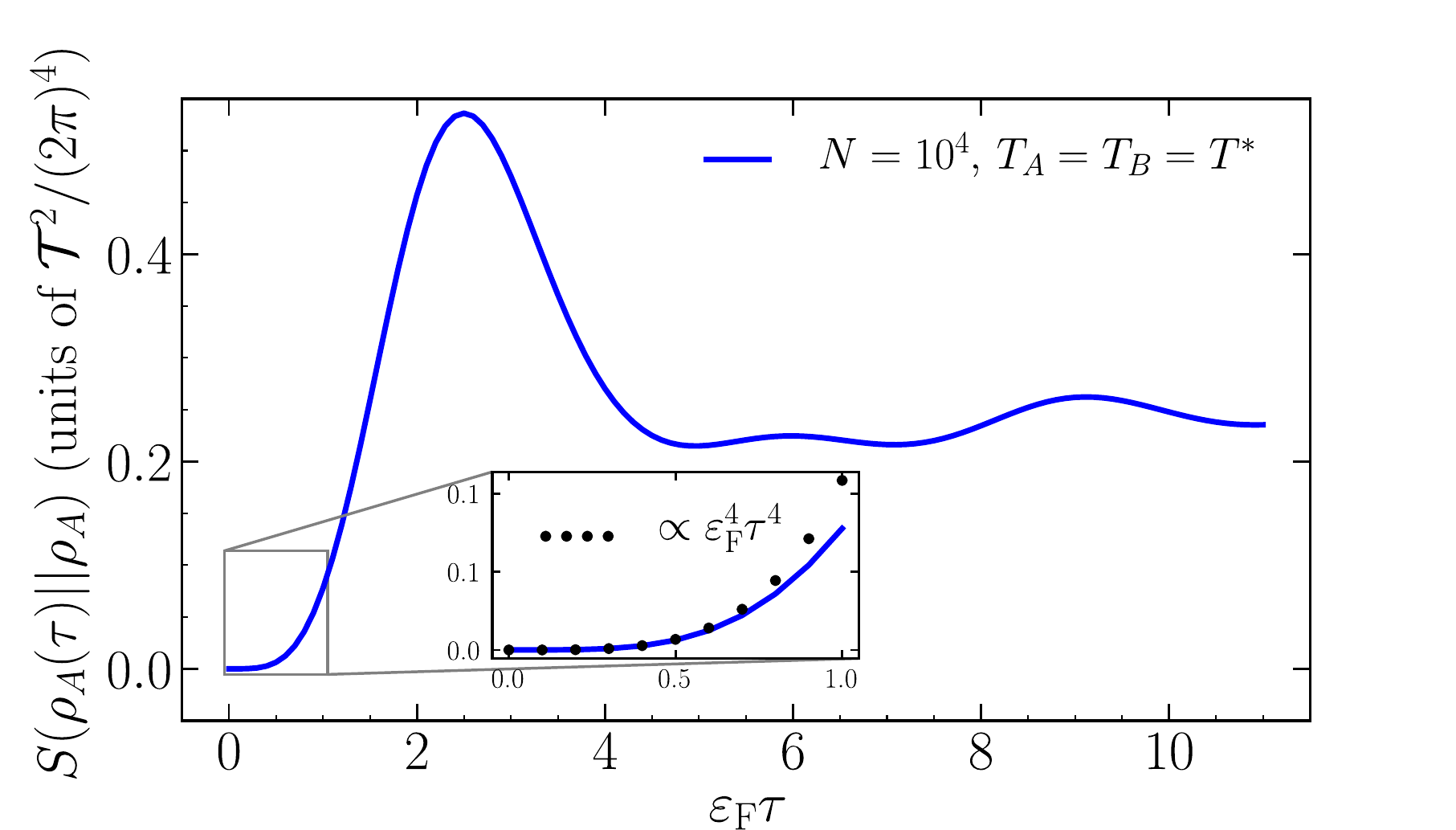} \vspace{-0.1cm}
\caption{\small \label{fig:S_rel} Relative entropy computed in the second order in transmission ${\cal T}$ at 
$T = T^*$. The black dots in the inset show the relative entropy within $\varepsilon_{\rm F} \tau \ll 1$ approximation.}
\end{figure}

Minding the low-temperature divergences, we aim to compute both the entropy production $\Delta S_{\rm vN}=S_{\rm vN}(\tau) - S_{\rm vN}^{(0)}$ and the relative entropy $S(\rho_A(\tau)||\rho_A)$ in the leading order in ${\cal T}$. We begin with evaluating the first correction to the occupation numbers analogously to the energy shift (\ref{dHdt}): $n^{(1)}_{\bf p}(\tau) = - \frac{|\lambda|^2}{V^2} \! \sum_{{\bf p}'}  \frac{\sin^2 (\delta\xi_{{\bf p p}'}\tau\!/2)}{(\delta\xi_{{\bf p p}'}\!/2)^2}  (n^{(0)}_{\bf p} \!- n^{(0)}_{{\bf p}'})$, where $\delta\xi_{{\bf p p}'} =\xi_{\bf p}-\xi_{{\bf p}'}$. Then we substitute $n^{(1)}_{\bf p}(\tau)$ into the first entropy contributions in Eqs.~(\ref{S_A_pert},\ref{S_rel_pert}). 
For temperatures in the range
\begin{align}
    T^* \lesssim T \ll \varepsilon_{\rm F}, \label{T_cond}
\end{align}
where $ T^*\!\sim\!\varepsilon_{\rm F}/\ln N$, we find the entropy production
\begin{align}
    \Delta S_{\rm vN} \simeq \,  \Delta E_A/T, \label{S_vN_pert}
\end{align}
where $\Delta E_A$ is the energy increment computed in the leading order in ${\cal T}$ in Eq.~(\ref{dHdt}). In turn,  the relative entropy is subleading in ${\cal T}$ compared to Eq.~(\ref{S_vN_pert}): 
\begin{align} 
    S(\rho_A(\tau)||\rho_A) =& \frac{2{\cal T}^2}{(2\pi)^4} \frac{\varepsilon_{\rm F}}{N} \!\int\limits_{-\varepsilon_{\rm F}}^{\varepsilon_{\rm F}}\!\!\!
    \!d\omega  
    J(\omega,\!\tau)^2 \cosh^2 \frac{\omega}{2T}, \label{S_rel_pert_2}
\end{align}
where $J(\omega,\!\tau) = \int_{-\varepsilon_{\rm F}}^{\varepsilon_{\rm F}} d\omega'\, \frac{\sin^2 (\delta\omega \tau / 2)}{(\delta\omega / 2)^2}  (n^{\!(0)} (\omega)\! -\! n^{\!(0)} (\omega'))$.

The lower temperature bound $T^*$ in Eq.~(\ref{T_cond}) extensively depends on the particle number in the reservoir and originates from regularizing the perturbative series for entropy that we discuss in the Appendix \ref{app:RE}. Consequently, within the settled temperature range (\ref{T_cond}), the perturbative expressions for the von Neumann entropy production (\ref{S_vN_pert}) and the relative entropy (\ref{S_rel_pert_2}) are well defined. Though the lower temperature bound is suppressed only logarithmically with $N$, for a trapped atomic cloud, e.g., see Ref. \cite{Krinner2014Observation}, $N\!\sim\!10^5$ atoms giving $T^*\!\sim\!0.1 \varepsilon_{\rm F}$ -- well within the experimental reach. 

The relative entropy (\ref{S_rel_pert_2}) is a measure of state separation indicating how far the evolved state of the reservoir is from its initial thermal state. In Fig.~\ref{fig:S_rel} we observe that the state separation occurs on a timescale of $\tau_{\rm F}\!\sim\!1/\varepsilon_{\rm F}$. For the initial temperatures of the reservoirs equal to $T^*$, this regime is well described by the $\varepsilon_{\rm F}\tau \ll 1$ approximation of Eq.~(\ref{S_rel_pert_2}), $ S(\rho_A(\tau)||\rho_A) \simeq 2 ({\cal T}^2/(2\pi)^4/N)  \varepsilon_{\rm F}^4 \tau^4 ( T \sinh (T/\varepsilon_{\rm F}) / \varepsilon_{\rm F} - 1 )$, illustrated in the inset in Fig.~\ref{fig:S_rel}. Alongside, 
the relative entropy is subleading to the energy contribution to the von Neumann entropy, which leaves us with the expression (\ref{S_vN_pert}) 
in the leading order in tunneling coefficient. Combining the perturbative result for the von Neumann entropy production (\ref{S_vN_pert}) and the energy balance in the reservoir upon decoupling (\ref{E_to_W}), we deduce that the von Neumann entropy of the reservoir $A$ accumulated during its mutual evolution with $B$ defines the work required to uncouple $A$ from $B$:
\begin{align}
    W \simeq 2 T \Delta S_{\rm vN}. \label{W}
\end{align}
The same is valid for the reservoir $B$ whose entropy equals the entropy of the reservoir $A$.


The work-to-entropy relation (\ref{W}) is derived for $T_A = T_B$ 
and, therefore, there is no heat flow. However, as seen from Eq.~(\ref{E_to_W}) and Fig.~\ref{fig:E_2}, even implying the initial temperature imbalance, the heat transfer is negligible compared to the work contribution to energy on a timescale of the Fermi time. 
Hence, on this timescale, one neglects the initial temperature imbalance and considers the 
 reservoirs at equal temperatures with no loss of generality. Combining Eqs.~(\ref{E_to_W},\ref{W}) with quantum-mechanical 
energy increment (\ref{dE_T0}), which defines the primary contribution to energy at low temperatures, we conclude that the von Neumann entropy is increasing on a timescale of $\tau_{\rm F}$, which is faster than the heat flow. 

Determination of the regime where the influence of the heat flow on the entropy dynamics can be discarded may allow one to characterize the quantum correlations' generation after the local quench at finite temperatures, in contrast with the previous studies strictly implying zero temperature \cite{Klich2009Quantum,Fradkin2009PRB}. Furthermore, despite being focused on the free homogeneous Fermi gas in the tunneling regime, we anticipate our results to hold beyond these limitations. Indeed, the relation (\ref{I_law}) is non-perturbative and does not depend on the microscopic details of the system. A predicted energy change associated with quench-induced entropy requires (a) the initial state of the reservoir described by a thermal state and (b) the quench potential that does not commute with the initial Hamiltonian of the system. Accordingly, in Ref. \cite{Ohanesjan2022Energy}, we study a setup with finite-size reservoirs with strongly-interacting fermions using exact diagonalization and confirm Eq.~(\ref{W}) on a timescale set by the inverse largest energy scale in the problem.


{\it Conclusion.---} Inspired by recent advances in ultracold Fermi gases, showing a drastic difference between the characteristic transport timescales in atomic systems compared to their electronic counterparts, 
we investigate the dynamics of Fermi gas on a timescale of $\tau_{\rm F}$ after driving the system out of equilibrium. 

We consider a two-terminal geometry confining two reservoirs filled with a cold Fermi gas and coupled by a tunneling contact. The reservoirs are initially at different temperatures, while the tunneling contact instantly opens at time $t=0$ and closes at $t=\tau$ by a series of two local quenches following one another. The energy change in either reservoir consists of the heat transferred from the hotter system to the colder one through the open tunneling contact and the work done by the second quench to decouple the reservoirs. For $\tau \sim \tau_{\rm F}$, we find that both reservoirs gain energy independent of their initial temperatures. This energy increment arises from work done by the second quench, whereas the heat 
contribution to the resulting energy change is negligible. We relate the work to the von Neumann entropy accumulated by the reservoir from $t=0$ to $t=\tau$. This relation grants 
a dynamic track of the von Neumann entropy production. 
On a timescale of $\tau_{\rm F}$, the quench-induced entropy production is positive 
and grows sufficiently faster than the heat flow. This provides a thermodynamic insight into the ultrafast out-of-equilibrium dynamics of Fermi gas and a possibility to characterize quantum correlations' generation at finite temperature. In turn, the $\tau_{\rm F}$ timeframe and the required temperature regime are experimentally accessible in ultracold atoms. 



{\it Acknowledgments.---} 
We have benefited from discussions with Boris Altshuler, Marcello Dalmonte, Rosario Fazio, Mikhail Kiselev, Anatoli Polkovnikov, Alessandro Silva, Yuxuan Wang, and Jan Zaanen. This research was supported in part by startup funds at the University of Florida, the Netherlands Organization for Scientific Research/Ministry of Science and Education (NWO/OCW), and by the European Research Council (ERC) under the European Union’s Horizon 2020 research and innovation programme. 

\appendix

\section{Derivation of the energy increment}\label{app:FGR}

In this Appendix we derive the energy increment formula (9) in the main text.  

\begin{figure*}[t!!]
\center \vspace{-0.28cm}
\includegraphics[width=0.443\linewidth]{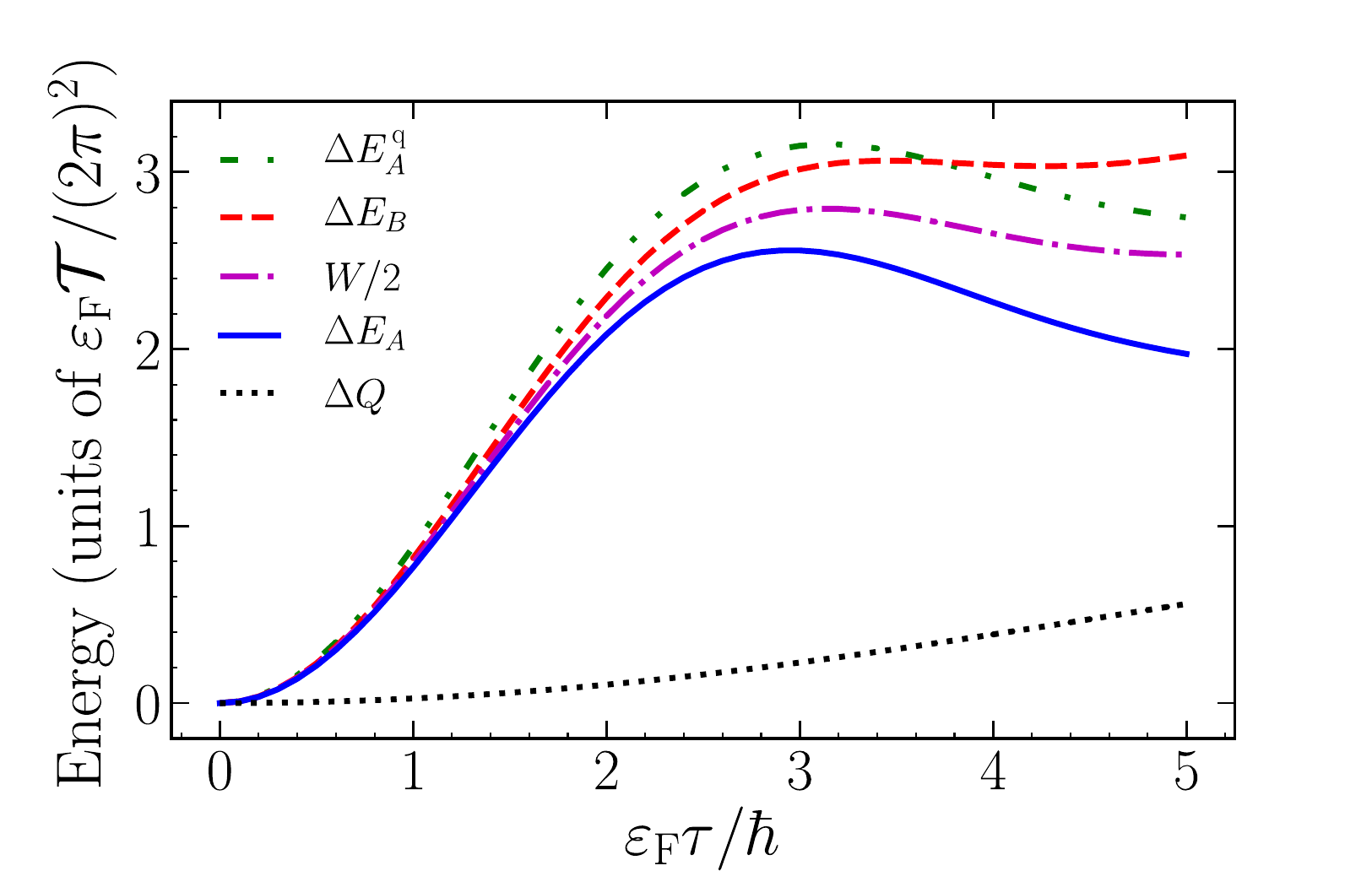} \qquad \qquad 
\includegraphics[width=0.443\linewidth]{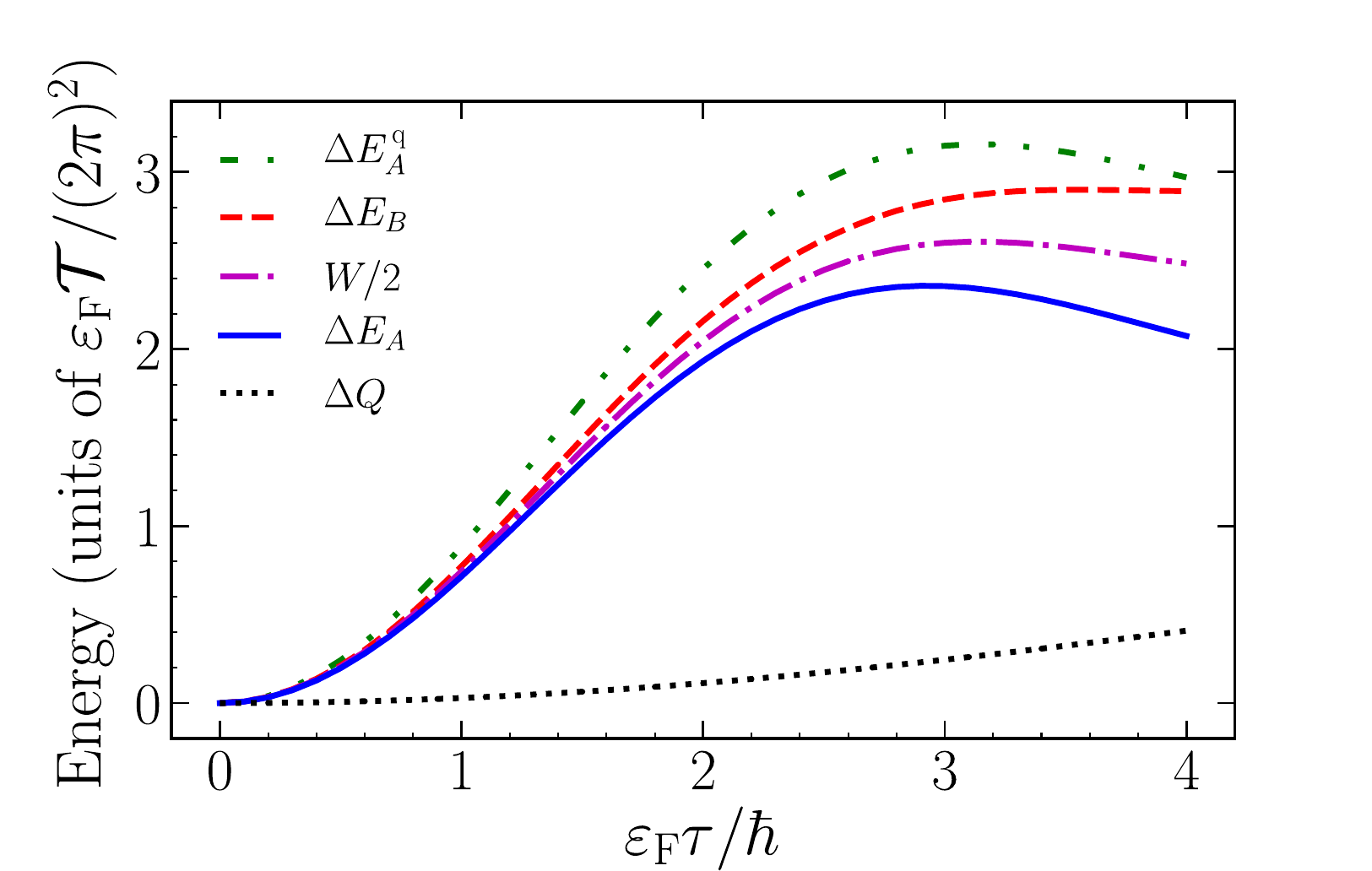} 
\caption{\small \label{fig:E_balance_hT} Energy balance in the system upon decoupling at $t=\tau$. $\Delta E_A^{\, \rm q}$ is the energy change at zero temperature ($T_A=T_B=0$), $\Delta E_B$ is the energy change in the initially colder system, $\Delta E_A$ is the energy change in the initially hotter system, $W$ is the work done by the quench upon decoupling, $\Delta Q$ is the heat transferred from $A$ to $B$. {\it Left panel}: $T_A = 0.2 T_{\rm F}$ and $T_B = 0.15 T_{\rm F}$. {\it Right panel}: $T_A = 0.25 T_{\rm F}$ and $T_B = 0.2 T_{\rm F}$. The Fermi temperature is defined as $T_{\rm F} = \varepsilon_{\rm F}/k_{\rm B}$. The maximum value on the time axis is chosen as $1/{\rm max}\lbrace T_A, T_B \rbrace$.}
\end{figure*}

To evaluate the overall energy shift in the reservoir $A$ after decoupling from $B$, we compute the corresponding energy flux
\begin{align} 
    \frac{d\langle H_A \rangle}{dt}  = - i g(t) \frac{1}{V}\sum_{{\bf p}{\bf p}'}\xi_{\bf p} ( \lambda \langle a^\dag_{\bf p} b_{{\bf p}'}\rangle -h.c. ), \label{app0:dHdt_def}
\end{align}
where $V$ is the size of the reservoir.
The correlation functions that define the energy flux (\ref{app0:dHdt_def}) satisfy the equation
\begin{align} \nonumber
    \lambda \frac{d\langle a_{\bf p}^\dag b_{{\bf p}'} \rangle}{dt}  =& i \lambda (\xi_{\bf p} - \xi_{{\bf p}'}) \langle a_{\bf p}^\dag b_{{\bf p}'}\rangle \\&- i g(t) \frac{|\lambda|^2}{V} \sum_{\bf q} (\langle a^\dag_{\bf p} a_{\bf q}\rangle - \langle b_{\bf q}^\dag b_{{\bf p}'} \rangle). \label{app0:eq_ab}
\end{align}
The exact solution of Eq.~(\ref{app0:eq_ab}) requires notion of the correlation functions $\langle a^\dag_{\bf p} a_{\bf q}\rangle$ and $\langle b_{\bf q}^\dag b_{{\bf p}'}\rangle$, the momenta-diagonal components of which are dynamic occupation numbers $n_{\alpha{\bf p}} (t)$ in the reservoirs $\alpha=A,B$. 
We solve Eq.~(\ref{app0:eq_ab}) perturbatively in the lowest order in $\lambda$ implying  the equilibrium occupation numbers of the initial state of the system. 
Since there are no inter-momenta couplings before the quench, we use the diagonal correlations $\langle a^\dag_{\bf p} a_{\bf q}\rangle = \delta_{{\bf p}{\bf q}}\langle a^\dag_{\bf p} a_{\bf p}\rangle_0 = \delta_{{\bf p}{\bf q}} n^{(0)}_{A{\bf p}}$ and  $\langle b^\dag_{\bf q} b_{{\bf p}'}\rangle = \delta_{{\bf q}{\bf p}'}\langle b^\dag_{{\bf p}'} b_{{\bf p}'}\rangle_0 = \delta_{{\bf q}{\bf p}'} n^{(0)}_B(\xi_{{\bf p}'})$ in  Eq.~(\ref{app0:eq_ab}), where $n^{(0)}_{\alpha {\bf p}}=\left(e^{\xi_{\bf p}/T_\alpha}+1\right)^{-1}$ is the Fermi distribution function. The sought-for correlation function is 
\begin{align} \nonumber
    \lambda \langle a_{\bf p}^\dag b_{{\bf p}'} \rangle =& - i\, \frac{|\lambda|^2}{V} (n^{(0)}_{A {\bf p}} - n^{(0)}_{B {\bf p}'}) \\ & \times  \int_0^t \!\! dt' e^{i(\xi_{\bf p}-\xi_{{\bf p}'})(t-t')} g(t'). \label{app0:ab}
\end{align}
It follows form the correlation function (\ref{app0:ab}) that $\langle V_{AB}(0)\rangle =0$, since there are no correlations between $A$ and $B$ at $t=0$.

Substituting Eq.~(\ref{app0:ab}) into Eq.~(\ref{app0:dHdt_def}), we derive the Fermi golden rule formula for the energy flux
\begin{align} \nonumber
    \frac{d\langle H_A \rangle}{dt} = & -  g(t) \, G \! \int \!\!d\omega  d\omega' \, \omega \,  \frac{\sin(\omega-\omega')t}{\pi(\omega-\omega')} \\&\times (n^{(0)}_A(\omega) - n^{(0)}_B(\omega')), \label{app0:dHdt}
\end{align}
where $G = 2\pi \nu_A \nu_B |\lambda|^2 = {\cal T}/(2\pi)$  is the particle conductance \footnote{Similar to Eq.~(\ref{app0:dHdt}), one can derive the particle current ${\cal I} = -   G \! \int \!\!d\omega  d\omega' \,  \frac{\sin(\omega-\omega')t}{\pi(\omega-\omega')}  (n^{(0)}_A(\omega) - n^{(0)}_B(\omega'))$. Taking the limit $t \to +\infty$ and implying the particle imbalance as $n^{(0)}_A(\omega) \to n^{(0)}_A(\omega+\mu)$, we get $d{\cal I}/d \mu = G$, where $G$ is the particle conductance \cite{Datta1995Electronic}. For $\hbar = 1$, the particle conductance has the same units as the transmission coefficient $\mathcal{T} = 2\pi G$, while restoration of the Plank's constant $h$ gives $G = \mathcal{T}/h$.} and $\nu_\alpha = V^{-1}\sum_{\bf p}\delta(\omega-\xi_{\bf p})$ is DoS for $\alpha=A,B$. Integrating the energy flux (\ref{app0:dHdt}) over time we obtain Eq.~(9) in the main text.

\section{Temperature corrections to the energy increment}\label{app:E_T}

In this Appendix, we compute finite temperature corrections to the energy increment (10) derived in the main text.

The temperature corrections to the zero-temperature energy increment (we imply ${\rm max}\lbrace T_A, T_B\rbrace \ll \varepsilon_{\rm F}$) can be conventionally computed by applying the Sommerfeld expansion to the energy integrals in Eq.~(9) in the main text:
\begin{align}
    \int_{-\varepsilon_{\rm F}}^{\varepsilon_{\rm F}} \!\!d\omega \frac{F(\omega, t)}{e^{\omega/T}+1} \simeq \int_{-\varepsilon_{\rm F}}^0 \!\! d\omega F(\omega, t) \!+\! \frac{\pi^2 T^2}{6} F'(0, t), \label{appA:Sommerfeld}
\end{align}
where $F'(0, t)$ denotes the energy derivative of an arbitrary function $F$ at $\omega=0$.
From Eq.~(9) in the main text we get 
\begin{align} \nonumber
    \Delta E_A =& - \frac{G}{\pi} \int_{0}^{\tau} dt \int_{-\varepsilon_{\rm F}}^{\varepsilon_{\rm F}}  d\omega \, ( F_A(\omega, t) n_A^{(0)}(\omega) \\ &- F_B(\omega, t) n_B^{(0)}(\omega) ), \label{appA:dE_T}
\end{align}
where
\begin{align}
F_A(\omega, t) &= \omega \int_{-\varepsilon_{\rm F}}^{\varepsilon_{\rm F}} d\omega' \,  \frac{\sin(\omega-\omega')t}{(\omega-\omega')}, \label{appA:FA} \\
F_B(\omega, t) &= \int_{-\varepsilon_{\rm F}}^{\varepsilon_{\rm F}} d\omega' \,  \omega'\,  \frac{\sin(\omega-\omega')t}{(\omega-\omega')}. \label{appA:Fb}
\end{align}

First, let's consider the cooling protocol for the reservoir $A$, so that $T_A = T \ll \varepsilon_{\rm F}$ and $T_B \ll T$. Using the equilibrium occupation numbers $n^{(0)}_A(\omega) = \left(e^{\omega/T}+1\right)^{-1}$ and $n^{(0)}_B(\omega)=\theta(-\omega)$, we apply the Sommerfeld expansion (\ref{appA:Sommerfeld}) to the first term in Eq.~(\ref{appA:dE_T}) and reinstate the result (10) in the main text with a small temperature correction:
\begin{align} 
    \Delta E_{A/B} \simeq &  \Delta E_A^{\, \rm q} \mp \frac{\cal T}{2\pi} \, \varepsilon_{\rm F}  \, \frac{\pi T^2}{3 \varepsilon_{\rm F}^2} \int_0^{\varepsilon_{\rm F} \tau} \!\! d\zeta  \int_0^{\zeta} \!\! d\zeta' \, \frac{\sin \zeta'}{\zeta'}, \label{appA:dE_T_cool}
\end{align}
where the minus/plus sign corresponds to the system $A$/$B$.
In the above, we used $F_A'(0, t) = 2\int^{\varepsilon_{\rm F} \tau}_0  d\zeta \, \sin \zeta/ \zeta$.

Similarly, for the reservoirs  prepared at the same temperature $T_A = T_B = T \ll \varepsilon_{\rm F}$ we obtain
\begin{align}
    \Delta E_A \simeq \Delta E_A^{\, \rm q} - \frac{\cal T}{2\pi} \, \varepsilon_{\rm F} \, \frac{4\pi T^2}{3 \varepsilon_{\rm F}^2} \sin^2\left(\frac{\varepsilon_{\rm F} \tau}{2}\right).
\end{align}

\section{Energy curves at higher temperature}\label{app:dQ}

In this Appendix, we show the energy balance similar to one depicted in Fig.~2 in the main text at higher temperatures. As demonstrated in Fig.~\ref{fig:E_balance_hT}, the energy increments in both reservoirs and the work can be considered equal and dominate the heat transfer on a timescale of $\sim 1/\varepsilon_{\rm F}$. Hence, they do not depend on the initial temperature imbalance in this regime.

\begin{figure}[b!!]
\center \vspace{-0.28cm}
\includegraphics[width=0.916\linewidth]{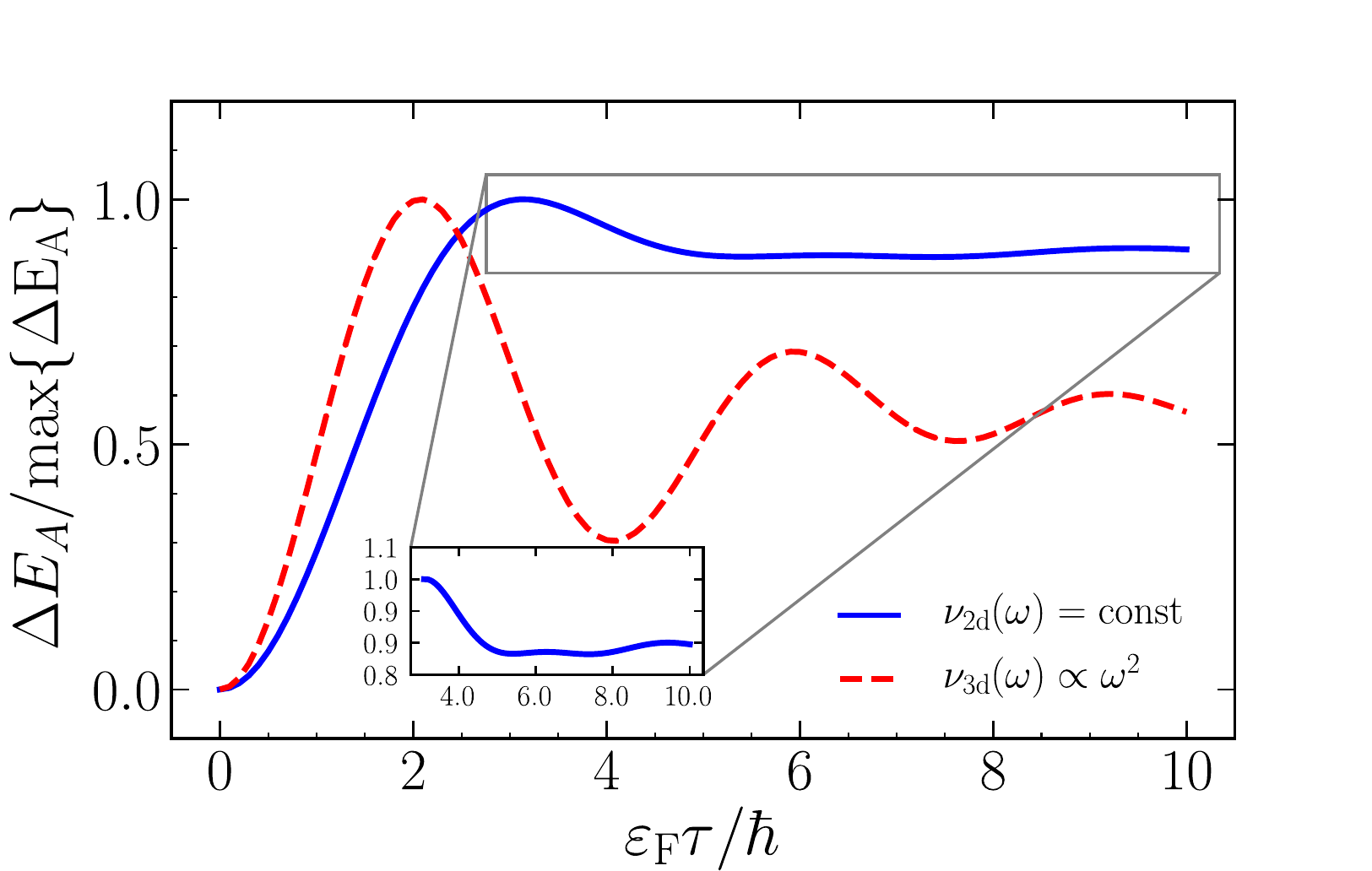}
\caption{\small \label{fig:dE_trap} Energy shift in the reservoir $A$ due to quench-coupling with $B$ for a $2d$ homogeneous Fermi gas ($\nu_{2d}(\omega)={\rm const}$) and for a $3d$ Fermi gas in a parabolic trap ($\nu_{3d}(\omega) \propto \omega^2$ \cite{Brantut2013Thermoelectric, Grenier2016Thermoelectric}). The reservoirs are assumed to be at equal initial temperatures $T_A = T_B = 0.1 \varepsilon_{\rm F}$. The energy curves are normalized by their maximal values. }
\end{figure}

\section{Energy curves for a harmonically trapped 3d Fermi gas}\label{app:trap}

In this Appendix we compare the energy curves obtained using the constant density of states (DoS) for a homogeneous $2d$ Fermi gas with the energy curves for a $3d$ Fermi gas in a harmonic trap considered in the recent experiment \cite{Brantut2013Thermoelectric}.

The density of states in atomic gases is usually energy-dependent due to the inhomogeneity of the trap potential. Here we consider a harmonically trapped three-dimensional Fermi gas with DoS $\nu_{3d}(\omega) \propto \omega^2$ \cite{Brantut2013Thermoelectric, Grenier2016Thermoelectric}. Analogously to Eq.~(9) in the main text, we evaluate the energy change in the reservoir from the Fermi golden rule formula, 
\begin{align} \nonumber
    \Delta E_A \propto& -
    \int_{-\varepsilon_{\rm F}}^{\varepsilon_{\rm F}} \!\! d\omega  d\omega' \, \omega^3 {\omega'}^2 \, \frac{\sin^2(\delta\omega\tau/2)}{ (\delta\omega/2)^2}  \\ & \times (n_A^{(0)}(\omega) - n^{(0)}_B(\omega')), \label{app:dE_trap}
\end{align} 
where we used $\nu_{3d}(\omega)\propto \omega^2$ and introduced $\delta \omega =\omega -\omega'$. In Fig.~\ref{fig:dE_trap}, we illustrate that the qualitative behaviour of the energy curves is similar for different energy dependencies of DoS. For times $\tau \lesssim 1/\varepsilon_{\rm F}$, both energy curves indicate power-law increase and subsequent oscillations and equilibration at later times. 

\section{Perturbation theory for entropy}\label{app:RE}

In this Appendix, we determine the temperature regime under which one can evaluate the von Neumann entropy within the quantum-mechanical perturbation theory. First, we regularize the low-temperature divergences for $\varepsilon_{\rm F}t\ll 1$. Then we show that the regularization is sufficient to compute the entropy at finite times in the leading order in ${\cal T}$.

We begin with expanding the von Neumann entropy in occupation numbers up to $l$-th order in transmission coefficient ${\cal T}$ and then splitting it into three contributions:
\begin{align}
    S_{\rm vN} =& S_{\rm vN}^{(0)} + \Delta S_{\rm vN},\\
    \Delta S_{\rm vN} =&  \frac{1}{T} \sum_{m=1}^l \sum_{\bf p} \xi_p n_{\bf p}^{(m)} - \sum_{m=1}^l S^{(m)}(\rho_A(t)||\rho_A). \label{appB:dS}
\end{align}
The first term in Eq.~(\ref{appB:dS}) is the energy contribution equal to $\Delta E_A/T$, the second one is given by the relative entropy generated by the perturbative series
\begin{align}
    S^{(2)}(\rho_A(t)||\rho_A) =& \sum_{\bf p} \frac{{n^{(1)}_{\bf p}}^2 }{2}  \bigg(\frac{1}{n^{(0)}_{\bf p}} + \frac{1}{1 -  n^{(0)}_{\bf p} }\!\bigg), \label{appB:S2_rel} \\ \nonumber
    S^{(3)}(\rho_A(t)||\rho_A) =& \sum_{\bf p} \! \bigg( n^{(1)}_{\bf p} n^{(2)}_{\bf p} \bigg(\frac{1}{n^{(0)}_{\bf p}} + \frac{1}{1 -  n^{(0)}_{\bf p}}\!\bigg) \\&-\!  \frac{{n^{(1)}_{\bf p}}^3 }{6}  \bigg(\frac{1}{{n^{(0)}_{\bf p}}^2} - \frac{1}{{(1 -  n^{(0)}_{\bf p})}^{\!2}} \!\bigg)\!\bigg), \label{appB:S3_rel} \\ \nonumber
    S^{(4)}(\rho_A(t)||\rho_A) =& \sum_{\bf p} \bigg(\! \bigg( \frac{{n^{(2)}_{\bf p}}^2}{2} + n^{(1)}_{\bf p} n^{(3)}_{\bf p} \bigg) \\ \nonumber &\times  \bigg(\frac{1}{n^{(0)}_{\bf p}} + \frac{1}{1 -  n^{(0)}_{\bf p} }\!\bigg) \\ \nonumber &-\!  \frac{{n^{(1)}_{\bf p}}^2 \! n^{(2)}_{\bf p} }{2}  \bigg(\frac{1}{{n^{(0)}_{\bf p}}^2} - \frac{1}{{(1 -  n^{(0)}_{\bf p})}^{\!2} }\!\bigg)\\ &+\!  \frac{{n^{(1)}_{\bf p}}^4}{12}  \bigg(\frac{1}{{n^{(0)}_{\bf p}}^3} + \frac{1}{{(1 -  n^{(0)}_{\bf p})}^{\!3} }\!\bigg) \!\bigg), \label{appB:S4_rel} \\ \nonumber  &\ldots ,
\end{align}
and $S^{(0)}_{\rm vN} = \pi^2 /3 \times N \times T/\varepsilon_{\rm F}$ is the initial value of the von Neumann entropy in the reservoir $A$. 
Here the superscript $m$ refers to the $m$-th order in ${\cal T}$.

The perturbative expansion of entropy contains low-temperature divergencies. The energy contributions to entropy given by $\Delta E_A/T$ diverge as $1/T$ for $T \to 0$ since $\Delta E_A$ does not depend on $T$ in this regime. As shown below, the relative entropy may diverge exponentially with temperature decrease in every order in ${\cal T}$.
Here our goal is to justify the leading in ${\cal T}$ contributions to the von Neumann entropy and the relative entropy by identifying and regularizing the low-temperature divergencies throughout the perturbative series. Though, we do not aim to compute the entropy in every order in perturbation theory. 

\subsection{Occupation numbers}\label{app:n}

At first, we analyze the occupation numbers in the leading order in ${\cal T}$ and $\varepsilon_{\rm F} t$.

Similar to Eqs.~(\ref{app0:dHdt_def}-\ref{app0:ab}) in Appendix \ref{app:FGR} we have
\begin{align} \nonumber
    \langle a_{\bf p}^\dag a_{{\bf k}} \rangle  =& \delta_{\bf p k} n_{\bf p}^{(0)} - \frac{i}{V} \int_0^t \!\! dt' e^{i (\xi_{\bf p} - \xi_{{\bf k}})(t-t')} g(t') \\& \times \sum_{{\bf p}'} (\lambda \langle a_{\bf p}^\dag b_{{\bf p}'}\rangle - \lambda^* \langle b_{{\bf p}'}^\dag a_{{\bf k}} \rangle), \label{appB:aa} 
\end{align}
and 
\begin{align} \nonumber
    \lambda \langle a_{\bf p}^\dag b_{{\bf p}'} \rangle  =& - i  \frac{|\lambda|^2}{V} \int_0^t \!\! dt' e^{i  (\xi_{\bf p} - \xi_{{\bf p}'})(t-t')} g(t') \\ &\times \sum_{\bf q} (\langle a^\dag_{\bf p} a_{\bf q}\rangle - \langle b_{\bf q}^\dag b_{{\bf p}'} \rangle). \label{appB:ab} 
\end{align}
Substituting Eq.~(\ref{appB:ab}) into Eq.~(\ref{appB:aa}) and looking for the leading contribution in $\varepsilon_{\rm F} t$, we obtain
\begin{align} \nonumber
    \langle a_{\bf p}^\dag a_{{\bf k}} \rangle =& \delta_{\bf p k} n_{\bf p}^{(0)} - \frac{|\lambda|^2}{V^2} \int_0^t \!\! dt' \! \int_0^{t'} \!\!\! dt'' \sum_{{\bf p}'{\bf q}} (\langle a^\dag_{\bf p} a_{\bf q}\rangle  \\&+ \langle a^\dag_{\bf q} a_{\bf k}\rangle - \langle b_{\bf q}^\dag b_{{\bf p}'}\rangle - \langle b_{{\bf p}'}^\dag b_{\bf q} \rangle ). \label{appB:eq_aa} 
\end{align}

Let's look for a perturbative solution of Eqs.~(\ref{appB:eq_aa}) in powers of $|\lambda|^2$: 
\begin{align}
    \langle a_{\bf p}^\dag a_{{\bf k}} \rangle =& \delta_{\bf p k} n_{\bf p}^{(0)} + \langle a_{\bf p}^\dag a_{{\bf k}} \rangle_1 + \ldots. 
\end{align}
For the first order in $|\lambda|^2$ we get 
\begin{align} \nonumber
    \langle a_{\bf p}^\dag a_{{\bf k}} \rangle_1 = & - \frac{|\lambda|^2 t^2}{2 V^2}  \sum_{{\bf p}'} (n_{\bf p}^{(0)} + n_{\bf k}^{(0)} - 2 n_{{\bf p}'}^{(0)}) \\ \nonumber =& 
     \frac{\nu |\lambda|^2 t^2}{2 V} \varepsilon_{\rm F}\!\left( \tanh\frac{\xi_{\bf p}}{2T} + \tanh\frac{\xi_{\bf k}}{2T}\right) \\ =& \langle b_{\bf k}^\dag b_{{\bf p}}\rangle_1, \label{appB:aa_1}
\end{align}
where we used $\sum_{\bf p} = \nu V \int_{-\varepsilon_{\rm F}}^{\varepsilon_{\rm F}}$ and assumed $T_A = T_B = T$. The diagonal component of Eq.~(\ref{appB:aa_1}) gives the correction to the occupation numbers:
\begin{align}
    n^{(1)}_{\bf p} = \frac{\cal T}{(2\pi)^2} \frac{(\varepsilon_{\rm F}t)^2}{N} \tanh \frac{\xi_{\bf p}}{2T}. \label{appB:n1}
\end{align}
Here we introduced the transmission coefficient ${\cal T} = (2 \pi)^2\nu^2 |\lambda|^2$ and the particle number $N = V p_{\rm F}^2/(4\pi) = V \nu \varepsilon_{\rm F}$. 

Similarly, in the next order in $|\lambda|^2$ 
\begin{align} \nonumber
    n^{(2)}_{\bf p} =& \frac{|\lambda|^4 t^4}{4! V^4} \sum_{{\bf p}'{\bf q p}''}\! \left.(n_{\bf p}^{(0)} + n_{\bf k}^{(0)} - 2 n_{{\bf p}''}^{(0)})\right|_{{\bf k}={\bf p}} \\=&- \frac{2{\cal T}^2}{(2\pi)^4} \frac{(\varepsilon_{\rm F}t)^4}{4!N} \tanh \frac{\xi_{\bf p}}{2T} . \label{appB:n2}
\end{align}

This procedure is easily generalized for arbitrary order:
\begin{align}
    n_{\bf p}^{(m)} = \frac{(-1)^{m+1}}{N} \frac{2{\cal T}^{m}}{(2\pi)^{2m}} \frac{(\varepsilon_{\rm F}t)^{2m}}{(2m)!} \tanh \frac{\xi_{\bf p}}{2T}, \label{appB:n_arb}
\end{align}
where the equation ({\ref{appB:n_arb}}) describes the leading in $\varepsilon_{\rm F} t$ contribution to the $m$-th order correction in $\cal T$ to the occupation numbers. 
However, formula ({\ref{appB:n_arb}}) is clearly inconsistent for an actual computation of the occupation numbers $n_{\bf p} = \sum_m n_{\bf p}^{(m)}$ since e.g. the terms $\propto {\cal T} (\varepsilon_{\rm F} t)^{2m}$ in $n_{\bf p}^{(1)}$ are neglected while the smaller terms $\propto {\cal T}^{m} (\varepsilon_{\rm F} t)^{2m}$ are kept. Thus, in a short-time regime, $\varepsilon_{\rm F} t \ll 1$, the accurate approximation for the occupation numbers is 
\begin{align}
    n_{\bf p} \simeq n_{\bf p}^{(0)} + n_{\bf p}^{(1)}
\end{align} 
found earlier in Eq.~(\ref{appB:n1}), whereas the terms $n_{\bf p}^{(m\geq2)}$ in Eq.~(\ref{appB:n_arb}) exceed the given accuracy. 
Nonetheless, Eq (\ref{appB:n_arb}) clarifies that the occupation numbers are $\propto N^{-1}$ and do not have any low-temperature divergences in any order in perturbation theory. 
These observations are essential for the relative entropy regularization.

\subsection{Entropy regularization}\label{app:regularize}

Now we proceed with regularizing the low-temperature divergences in the entropy perturbative expansion.

The energy term in Eq.~(\ref{appB:dS}) diverges only as $1/T$ for $T \to 0$ since the occupation numbers are regular in all orders in perturbation theory. Hence, we imply $|\Delta E_A|\ll T |S_{\rm vN}^{(0)}|$ to regularize the divergence. In a short-time limit, $\Delta E_A \simeq \sum_{\bf p} \xi_{\bf p} n_{\bf p}^{(1)} = {\cal T}/(2\pi)^2 \times \varepsilon_{\rm F} \times (\varepsilon_{\rm F} t)^2$ that when compared with the initial value of the von Neumann entropy $S^{(0)}_{\rm vN} = \pi^2 /3 \times N \times T/\varepsilon_{\rm F}$ leads to
\begin{align}
    \frac{\cal T}{(2\pi)^2} (\varepsilon_{\rm F} t)^2 \times \frac{3}{\pi^2 N} \ll \frac{T^2}{\varepsilon_{\rm F}^2}. \label{appB:cond0}
\end{align}
For ${\cal T}/(2\pi)^2\ll 1$ and $\varepsilon_{\rm F} t \ll 1$ we replace the condition (\ref{appB:cond0}) with $\sqrt{{\cal T}} \varepsilon_{\rm F} t /(2\pi) \,  \, T^* \ll T^* \lesssim T$, which is fulfilled for $T^* \lesssim T$:
\begin{align}
    \frac{T^*}{\varepsilon_{\rm F}} = \sqrt{\frac{3}{\pi^2 N}} . \label{appB:cond1}
\end{align}
By excluding ${\cal T}/(2\pi)^2$ and $\varepsilon_{\rm F} t$ from inequality (\ref{appB:cond0}) and in the regularization conditions below, we ensure that, when considering the higher-order terms in the entropy expansion, every next order term is smaller than the previous one by ${\cal T}/(2\pi)^2 \times (\varepsilon_{\rm F} t)^2$, the same as it happens for the occupation numbers and energy corrections. 

The relative entropy is subleading in ${\cal T}$  to the energy contribution in Eq.~(\ref{appB:dS}) but contains exponential divergences at low temperatures. Therefore, to stabilize the perturbative expansion we require $T |S^{(2)}(\rho_A(t)||\rho_A)|\ll |\Delta E_A|$.
Substituting the occupation numbers from Eq.~(\ref{appB:n1}) into the first finite contribution to the relative entropy in Eq.~(\ref{appB:S2_rel}) we find
\begin{align}
    S^{(2)}(\rho_A(t)||\rho_A) \simeq  \frac{2{\cal T}^2}{(2\pi)^4} \! \frac{(\varepsilon_{\rm F} t)^4}{N} \! \left( \frac{T}{\varepsilon_{\rm F}}\sinh \frac{\varepsilon_{\rm F}}{T} - 1 \!\right)\! \label{appB:S_2_st}
\end{align}
that explicitly depends on the particle number $N$.

\begin{figure}[t!!]
\center \vspace{-0.28cm}
\includegraphics[width=0.916\linewidth]{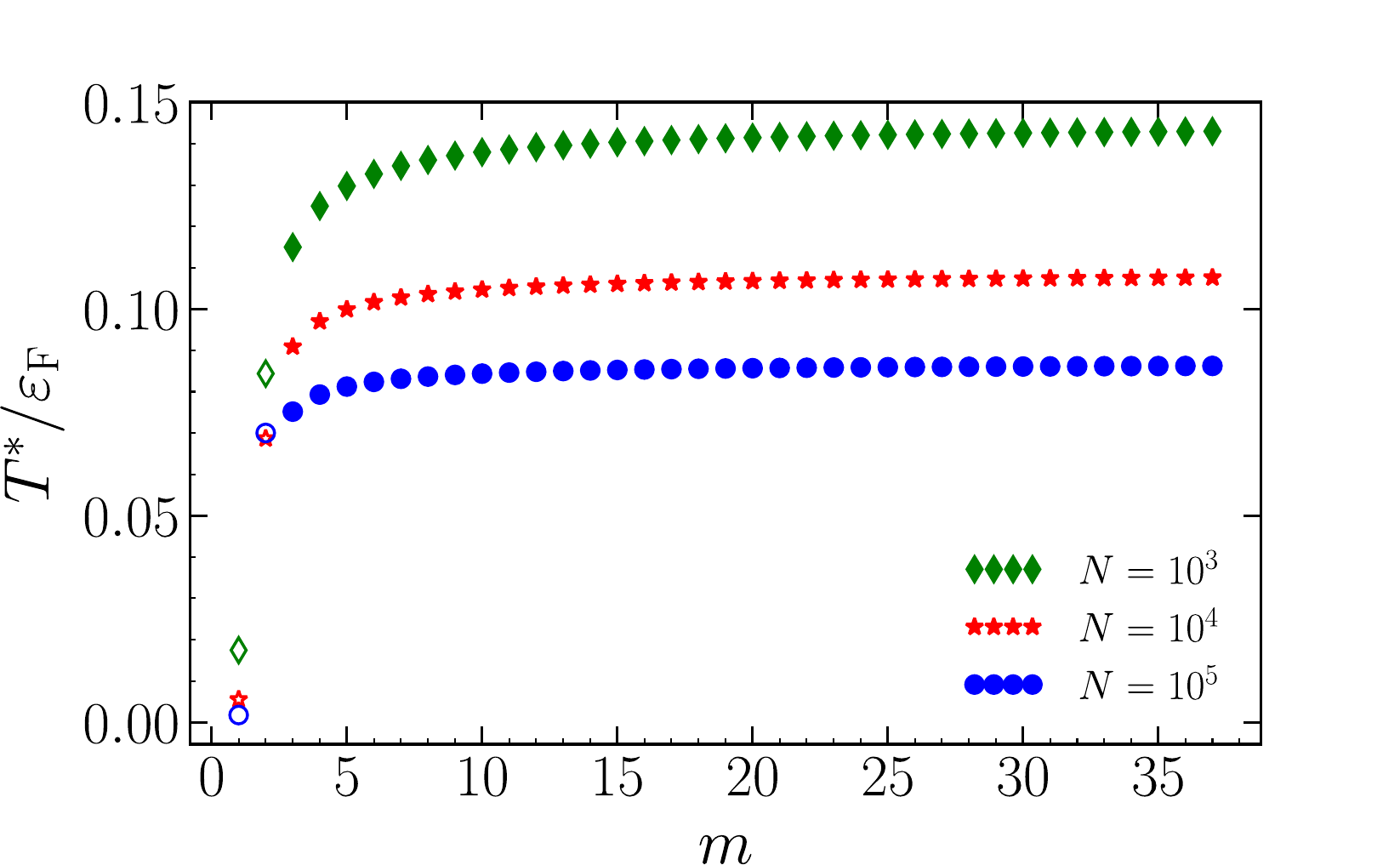}
\caption{\small \label{fig:T_m} Lower bound on temperature $T^*$ as a function of the order $m$ of the perturbative expansion for the relative entropy for a given particle number $N$. The first two ``punctured'' points for each plot are defined by Eqs.~(\ref{appB:cond1},\ref{appB:cond2}) correspondingly.}
\end{figure}

Alike the previous condition (\ref{appB:cond1}), if ${\cal T}/(2\pi)^2\ll 1$ and $\varepsilon_{\rm F} t \ll 1$, the requirement $T |S^{(2)}(\rho_A(t)||\rho_A)|\ll |\Delta E_A|$ will be satisfied for temperatures $T^* \lesssim T$, where $T^*$ is found from
\begin{align}
    \frac{T^*}{\varepsilon_{\rm F}}\left|\frac{T^*}{\varepsilon_{\rm F}}\sinh  \frac{\varepsilon_{\rm F}}{T^*} - 1\right| = \frac{N}{2}, \label{appB:cond2}
\end{align}
so that the large $N$ balances out the exponential growth for temperatures $T^* \lesssim T$. Overall, this restriction sufficiently pushes the lower temperature bound $T^*$, found earlier in Eq.~(\ref{appB:cond1}), towards the Fermi temperature, as shown in Fig.~\ref{fig:T_m}. To check if continuing the regularization pushes out the lower temperature bound above the Fermi temperature, which would break the requirement $T \ll \varepsilon_{\rm F}$, or whether $T^*$ converges to some value below $\varepsilon_{\rm F}$ we proceed with our analysis.

Now let's consider the relative entropy order by order. 
From Eqs.~(\ref{appB:S3_rel},\ref{appB:n2}) we notice that the first term in $S^{(3)}(\rho_A(t)||\rho_A)$ has the same temperature and $N$ dependencies as $S^{(2)}(\rho_A(t)||\rho_A)$, where the later one has been already regularized for $T^* \lesssim T$. 
Withal, in the short-time limit the first term in $S^{(3)}(\rho_A(t)||\rho_A)$ is $\propto {\cal T}^3 t^6$ while $S^{(2)}(\rho_A(t)||\rho_A) \propto {\cal T}^2 t^4$. As such, the first term in $S^{(3)}(\rho_A(t)||\rho_A)$ clearly exceeds the accuracy in $\varepsilon_{\rm F} t$ since the larger contribution $\propto {\cal T}^2 t^6$ has been already neglected in the previous order. Therefore, as we aim for the leading order in $\varepsilon_{\rm F} t$, the first term in Eq.~(\ref{appB:S3_rel}) should be omitted. 

The same logic applies to the higher-order terms. For example, the first two terms in $S^{(4)}(\rho_A(t)||\rho_A)$ turn out to be redundant when compared to $S^{(2)}(\rho_A(t)||\rho_A)$ and $S^{(3)}(\rho_A(t)||\rho_A)$ assuming that the last term in $S^{(3)}(\rho_A(t)||\rho_A)$ is regular. Importantly, the omitted terms do not contain any extra low-temperature divergences since the occupation numbers are regular in any order in perturbation theory.

Finally, we address the remaining terms in the relative entropy that contain ${n_{\bf p}^{(1)}}^{m}$ in $m$-th order in perturbation theory and stabilize the entropy expansion. Regularizing these terms is crucial since they bring a new low-temperature divergence at every order in perturbation theory. 
To do so we require 
$|\widetilde{S}^{(m)}(\rho_A(t)||\rho_A)| \ll |\widetilde{S}^{(m-1)}(\rho_A(t)||\rho_A)|$ for all $m \geq 3$, where
\begin{align} \nonumber
    \widetilde{S}^{(m)}(\rho_A(t)||\rho_A) =&   \sum_{\bf p} \frac{{n^{(1)}_{\bf p}}^m }{m(m-1)}   \\ \nonumber & \times \bigg(\frac{(-1)^m}{{n^{(0)}_{\bf p}}^{m-1}} + \frac{1}{{(1 -  n^{(0)}_{\bf p})}^{\!m-1}} \!\bigg) \\  =& \frac{2{\cal T}^{m}(\varepsilon_{\rm F} t)^{2m}}{(2\pi)^{2m} N^{m-1}} \, \frac{T}{\varepsilon_{\rm F}} \, s_m(T/\varepsilon_{\rm F})  \label{appB:S_m_rel}
\end{align} 
and 
\begin{align} \nonumber
    s_m(x) =&  \frac{1}{m(m-1)} \bigg(\frac{1}{x} + \sum_{k=1}^{m} \frac{m!}{k!(m-k)!} \\  &\times \bigg(-\frac{1}{x} + \sum_{j=1}^{k-1} (-1)^{j+1} \, \frac{2}{j}\sinh\frac{j}{x} \bigg) \bigg). \label{appB:s_m}
\end{align}

\begin{figure}[t!!]
\center \vspace{-0.28cm}
\includegraphics[width=0.916\linewidth]{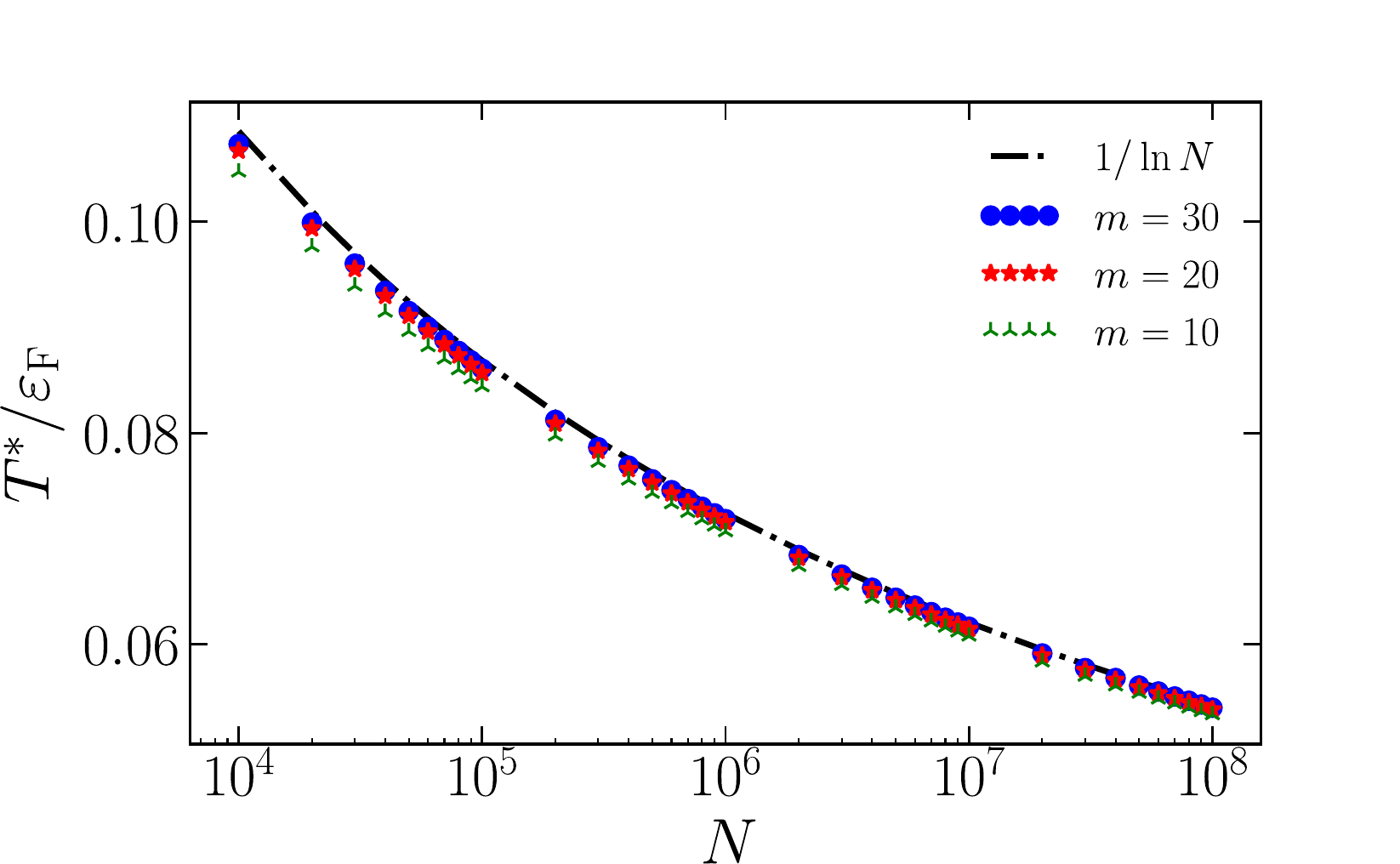} 
\caption{\small \label{fig:T_N} Lower bound on temperature $T^*$ as a function of the particle number $N$.}
\end{figure}

As in the case of the two previous conditions stated in Eqs.~(\ref{appB:cond1},\ref{appB:cond2}), supposed ${\cal T}/(2\pi)^2 \ll 1$ and $\varepsilon_{\rm F} t \ll 1$, one can determine the temperature regime $T^* \lesssim T$ that provides $|\widetilde{S}^{(m)}(\rho_A(t)||\rho_A)| \ll |\widetilde{S}^{(m-1)}(\rho_A(t)||\rho_A)|$. Using Eq.~(\ref{appB:S_m_rel}), we find the temperature $T^*$ from
\begin{align}
    |s_m(T^*/\varepsilon_{\rm F})| = N |s_{m-1}(T^*/\varepsilon_{\rm F})| \label{appB:cond3}
\end{align}
at every $m \geq 3$. In Fig.~\ref{fig:T_m} we show that $T^*$ computed from Eq.~(\ref{appB:cond3}) converges to a saturated value for $m \gg 3$. In its turn, the saturated value of the lower temperature bound is given by $T^*\!\sim\!\varepsilon_{\rm F}/\ln N$ as demonstrated in Fig.~\ref{fig:T_N}.
Now, when $\varepsilon_{\rm F}/\ln N \lesssim T$, every next order term in the relative entropy is regular in temperature and smaller than the previous one by ${\cal T}/(2\pi)^2 \times (\varepsilon_{\rm F} t)^2 \ll 1$. This makes $S^{(m\geq 3)}(\rho_A(t)||\rho_A)$ rightfully excessive in comparison to $S^{(2)}(\rho_A(t)||\rho_A)$.

Thereby, we have shown that for temperatures 
\begin{align}
    T^* \lesssim T \ll \varepsilon_{\rm F}, \label{appB:cond4}
\end{align}
where $ T^*\!\sim\!\varepsilon_{\rm F}/\ln N$,
the leading in ${\cal T}$ and $\varepsilon_{\rm F} t$ contributions to the von Neumann entropy and to the relative entropy are 
\begin{align}
    \Delta S_{\rm vN} \simeq & \frac{\cal T}{(2\pi)^2} \, \frac{\varepsilon_{\rm F}}{T} \, (\varepsilon_{\rm F} t)^2 \label{appB:S_vN_st} 
\end{align}
and
\begin{align}    
    S(\rho_A(t)||\rho_A) \simeq & \frac{2{\cal T}^2}{(2\pi)^4} \! \frac{(\varepsilon_{\rm F} t)^4}{N} \! \left( \frac{T}{\varepsilon_{\rm F}}\sinh \frac{\varepsilon_{\rm F}}{T} - 1 \!\right)\!. \label{appB:S_rel_st}
\end{align}

\subsection{Finite time}\label{app:finite_time}

Having regularised the entropy in a short-time approximation ($\varepsilon_{\rm F} t \ll 1$), we check if the regularization condition (\ref{appB:cond4}) is sufficient to compute the entropy on a timescale of $1/\varepsilon_{\rm F}$.

Substituting the first correction to the occupation numbers 
\begin{align} \nonumber
    n^{(1)}_{\bf p}(t) =\! - \frac{|\lambda|^2}{V^2} \sum_{{\bf p}'}  \frac{\sin^2 (\delta\xi_{{\bf p p}'} t/2)}{(\delta\xi_{{\bf p p}'}/2)^2}  (n^{(0)}_{\bf p} \!- n^{(0)}_{{\bf p}'})
\end{align}
with $\delta\xi_{{\bf p p}'} =\xi_{\bf p}-\xi_{{\bf p}'}$ into the lowest order in ${\cal T}=(2\pi)^2\nu^2 |\lambda|^2$ contribution to the relative entropy (\ref{appB:S2_rel}), we get
\begin{align} 
    S(\rho_A(t)||\rho_A)\! =& \frac{2{\cal T}^2}{(2\pi)^4} \frac{\varepsilon_{\rm F}}{N} \int_{-\varepsilon_{\rm F}}^{\varepsilon_{\rm F}}\!\!\! d\omega  J(\omega, \!t)^2 \cosh^2 \!\frac{\omega}{2T}, \label{appB:S_rel} \\  
    J(\omega, \!t)\! =&  t^2 \!\!\!  \int_{-\varepsilon_{\rm F}}^{\varepsilon_{\rm F}} \!\!\! d\omega'  F(\delta \omega t)  (n^{\!(0)}\!(\omega)\! -\! n^{\!(0)}\!(\omega')),\label{appB:J} \\
    F(\delta \omega t) =& \frac{\sin^2(\delta \omega t/2)}{(\delta\omega t/2)^2},\label{appB:Filter}
\end{align}
where we used that $\sum_{\bf p} = \nu V \int_{-\varepsilon_{\rm F}}^{\varepsilon_{\rm F}}\!\! d\omega$ and introduced $\delta\omega=\omega-\omega'$. 
In a short-time approximation, $F(\delta \omega t)\to 1$ that leads to our earlier result (\ref{appB:S_rel_st}). At finite time, $F(\delta \omega t)$ works as a spectral filter that suppresses the energy integral in Eq.~(\ref{appB:S_rel}). Once we regularize the short-time dynamics imposing the temperature restriction (\ref{appB:cond4}), the finite time regime will be taken care of by $F(\delta \omega t)$. Since the relative entropy is maximum at the lowest temperature, we plot the relative entropy (\ref{appB:S_rel}) at $T = T^*$ in Fig.~3 in the main text. From there we observe that the relative entropy is free of low-temperature divergences insofar as $\left.S(\rho_A(t\!\sim\!1/\varepsilon_{\rm F})||\rho_A)\right|_{T=T^*} \lesssim {\cal T}^2/(2\pi)^4 \times {\cal O}(1)$. The same is expected in the higher order contributions to the relative entropy so that they remain subleading in ${\cal T}$. In turn, the relative entropy computed in a short-time limit (\ref{appB:S_rel_st}) well approximates Eq.~(\ref{appB:S_rel}) for $t \lesssim 1/\varepsilon_{\rm F}$.

\bibliography{refs} 

\end{document}